\documentclass[aps,twocolumn,superscriptaddress]{revtex4}

\usepackage[dvips]{graphicx}

\newcommand{ \rts }{ \sqrt{s_{_{\rm NN}}} }

\newcommand{ \pbp }{ \overline{p}/p }

\def \pbar  {\overline{p}}

\def \hij   {HIJING}
\def \mpt   {$\langle p_t \rangle$}

\def \mbeta {$\langle \beta_t \rangle$}
\begin{document}
\title{Rapidity and Centrality Dependence of Proton and Anti-proton
  Production from $^{197}$Au+$^{197}$Au Collisions at $\rts = 130$~GeV
  }

\affiliation{Argonne National Laboratory, Argonne, Illinois 60439}
\affiliation{Brookhaven National Laboratory, Upton, New York 11973}
\affiliation{University of Birmingham, Birmingham, United Kingdom}
\affiliation{University of California, Berkeley, California 94720}
\affiliation{University of California, Davis, California 95616}
\affiliation{University of California, Los Angeles, California 90095}
\affiliation{Carnegie Mellon University, Pittsburgh, Pennsylvania 15213}
\affiliation{Creighton University, Omaha, Nebraska 68178}
\affiliation{Nuclear Physics Institute AS CR, \v{R}e\v{z}/Prague, Czech Republic}
\affiliation{Laboratory for High Energy (JINR), Dubna, Russia}
\affiliation{Particle Physics Laboratory (JINR), Dubna, Russia}
\affiliation{University of Frankfurt, Frankfurt, Germany}
\affiliation{Indiana University, Bloomington, Indiana 47408}
\affiliation{Insitute  of Physics, Bhubaneswar 751005, India}
\affiliation{Institut de Recherches Subatomiques, Strasbourg, France}
\affiliation{University of Jammu, Jammu 180001, India}
\affiliation{Kent State University, Kent, Ohio 44242}
\affiliation{Lawrence Berkeley National Laboratory, Berkeley, California 94720}\affiliation{Max-Planck-Institut fuer Physik, Munich, Germany}
\affiliation{Michigan State University, East Lansing, Michigan 48824}
\affiliation{Moscow Engineering Physics Institute, Moscow Russia}
\affiliation{City College of New York, New York City, New York 10031}
\affiliation{NIKHEF, Amsterdam, The Netherlands}
\affiliation{Ohio State University, Columbus, Ohio 43210}
\affiliation{Panjab University, Chandigarh 160014, India}
\affiliation{Pennsylvania State University, University Park, Pennsylvania 16802}
\affiliation{Institute of High Energy Physics, Protvino, Russia}
\affiliation{Purdue University, West Lafayette, Indiana 47907}
\affiliation{University of Rajasthan, Jaipur 302004, India}
\affiliation{Rice University, Houston, Texas 77251}
\affiliation{Universidade de Sao Paulo, Sao Paulo, Brazil}
\affiliation{University of Science \& Technology of China, Anhui 230027, China}
\affiliation{Shanghai Institute of Nuclear Research, Shanghai 201800, P.R. China}
\affiliation{SUBATECH, Nantes, France}
\affiliation{Texas A \& M, College Station, Texas 77843}
\affiliation{University of Texas, Austin, Texas 78712}
\affiliation{Valparaiso University, Valparaiso, Indiana 46383}
\affiliation{Variable Energy Cyclotron Centre, Kolkata 700064, India}
\affiliation{Warsaw University of Technology, Warsaw, Poland}
\affiliation{University of Washington, Seattle, Washington 98195}
\affiliation{Wayne State University, Detroit, Michigan 48201}
\affiliation{Institute of Particle Physics, CCNU (HZNU), Wuhan, 430079 China}
\affiliation{Yale University, New Haven, Connecticut 06520}

\author{J.~Adams}\affiliation{University of Birmingham, Birmingham, United Kingdom}
\author{C.~Adler}\affiliation{University of Frankfurt, Frankfurt, Germany}
\author{Aggarwal}\affiliation{Panjab University, Chandigarh, 160014, India}
\author{Z.~Ahammed}\affiliation{Purdue University, West Lafayette, Indiana 47907}
\author{J.~Amonett}\affiliation{Kent State University, Kent, Ohio 44242}
\author{B.D.~Anderson}\affiliation{Kent State University, Kent, Ohio 44242}
\author{M.~Anderson}\affiliation{University of California, Davis, California 95616}
\author{D.~Arkhipkin}\affiliation{Particle Physics Laboratory (JINR), Dubna, Russia}
\author{G.S.~Averichev}\affiliation{Laboratory for High Energy (JINR), Dubna, Russia}
\author{S.K.~Badyal}\affiliation{University of Jammu, Jammu, 180001, India}
% ...

\date{\today}

\begin{abstract}
  We report on the rapidity and centrality dependence of proton and
  anti-proton transverse mass distributions from
  $^{197}$Au+$^{197}$Au collisions at $\rts = 130$~GeV as measured by
  the STAR experiment at RHIC.  Our results are from the rapidity and
  transverse momentum range of $|y|<0.5$ and 0.35 $<p_t<$1.00~GeV/$c$.
  For both protons and anti-protons, transverse mass distributions
  become more convex from peripheral to central collisions
  demonstrating characteristics of collective expansion. The measured
  rapidity distributions and the mean transverse momenta versus rapidity 
  are flat within $|y|<0.5$. Comparisons of our data with 
  results from model calculations indicate that 
  in order to obtain a consistent picture of the proton(anti-proton)
  yields and transverse mass distributions the possibility of
  pre-hadronic collective expansion may have to be taken into account.
\end{abstract}


\maketitle

High energy nuclear collisions provide a unique opportunity to study
matter under extreme conditions for which one expects the formation of
a system dominated by deconfined quarks and gluons \cite{qm01}.  In
the search for this deconfined state, baryons play an important role.
Incoming beam baryons provide the energy for particle production and
development of collective motion.  It has systematically been observed
that the net-baryon number determines the chemical properties
\cite{pbm02}. In addition, baryon transport and baryon production
during the collision are particularly interesting because of their
dynamical nature
\cite{buza88,hansen95,e814proton,e866pbarp,na44proton,na49netproton}.
However, these are difficult processes due to their non-perturbative
features \cite{date85,kapusta01}.  At the RHIC energy $\rts$ =130~GeV,
anti-proton to proton ratios and yields at mid-rapidity have been
reported by several
experiments~\cite{starpbarp,phenixpikap,bramhs02,phobos_ratio}.  In
the region of $p_t \sim$ 2-3~GeV/$c$, the yield of protons approaches
that of pions \cite{phenixpikap} in central collisions. The exact
origin of this behavior is not clear and systematic measurements of
baryon distributions are important.

In this Letter, we present a systematic measurement of proton and
anti-proton production in Au+Au collisions at $\sqrt{s_{NN}}=130$ GeV
in the rapidity range $-0.5 < y < 0.5$ and for transverse momenta
$0.35 < p_t < 1.00$ GeV/c.  In particular, we report the first RHIC
measurements of the rapidity dependence of the proton and anti-proton yields,
essential for exploring the existence of a boost-invariant region in
the system.  We also study the centrality dependence of the yields and
mean transverse momenta for protons and anti-protons.  These results
allow for a detailed comparison to model predictions of proton and anti-proton
production at RHIC.

Two independent $^{197}$Au beams with an energy of 65~GeV per nucleon
were provided by the Relativistic Heavy Ion Collider~(RHIC) at the
Brookhaven National Laboratory. These beams collided around the
geometric center of the Solenoid Tracker at RHIC (STAR).  
Charged particles stemming from these collisions were measured 
in a large volume Time Projection Chamber (TPC) \cite{tpc}.  
A large solenoidal magnet of 0.25~T field strength provided 
momentum dispersion in the direction transverse to the beam line. 

For this analysis, we used 320k events with a minimum bias trigger and
154k events with a trigger selecting the 10\% most central events~\cite{starpbarp}.
Events with a primary vertex within $\pm30$~cm of the geometric center
of the TPC along the beam axis were accepted.  Tracks were required to
have at least 23 out of 45 maximum possible space points in the TPC
and to extrapolate back to the primary vertex within 2~cm (distance of
closest approach, $dca$).  To define the collision centrality, the
measured raw multiplicity distribution of charged particles within the
pseudorapidity range $| \eta |<0.75$ was divided into eight bins. The
highest centrality bin corresponds to 6\% of the measured cross
section for $^{197}$Au+$^{197}$Au collisions~\cite{star_v2}.
Protons and anti-protons were identified by correlating their energy
loss $dE/dx$ due to ionization in the TPC gas with the measured
momentum. This method has already been presented
in~\cite{starpbarp}.

The track reconstruction efficiency was determined by embedding simulated tracks 
into real events at the raw data level and subsequently applying the full reconstruction
algorithm to those events. The propagation of single tracks was performed
using the GEANT Monte Carlo code with a detailed model of the STAR
geometry and a realistic simulation of the TPC response. 
The resulting track reconstruction efficiency is $>70$\% at $p_t >$~0.5~GeV/$c$ for all
centralities. By varying the track cuts, the overall systematic uncertainty 
in the track reconstruction efficiency is estimated to be less than 10\%. 
Further, the relative resolution in transverse momentum 
was derived to be $\approx4\%$ at $p_t$ = 0.5~GeV/$c$.

Secondary interactions of particles with the detector material
generated background protons. Due to their different geometric
origin, these background protons appear as a rather flat tail in the
$dca$-distribution which extends into the peak region of primary
protons at small $dca$. In order to correct for
background protons, the proton $dca$--distribution was fitted by the
scaled anti-proton $dca$--distribution (which is background free) plus
the results on the proton background from Monte Carlo calculations.
Raw yields were extracted for protons and anti-protons with
$dca<2.0$~cm, optimizing the signal to background ratio for protons.
The raw yields were then corrected for track reconstruction
efficiency, proton background and in the case of anti-protons,
for absorption in the detector material.
The detector acceptance for  protons(anti-protons) from 
the decay of lambdas(anti-lambdas) or other
hyperons(anti-hyperons) is estimated to be larger than 95\%.
Corrections for feeddown from decays of hyperons(anti-hyperons) were not applied.
%
\begin{figure}[tbh]
\centering\mbox{
\includegraphics{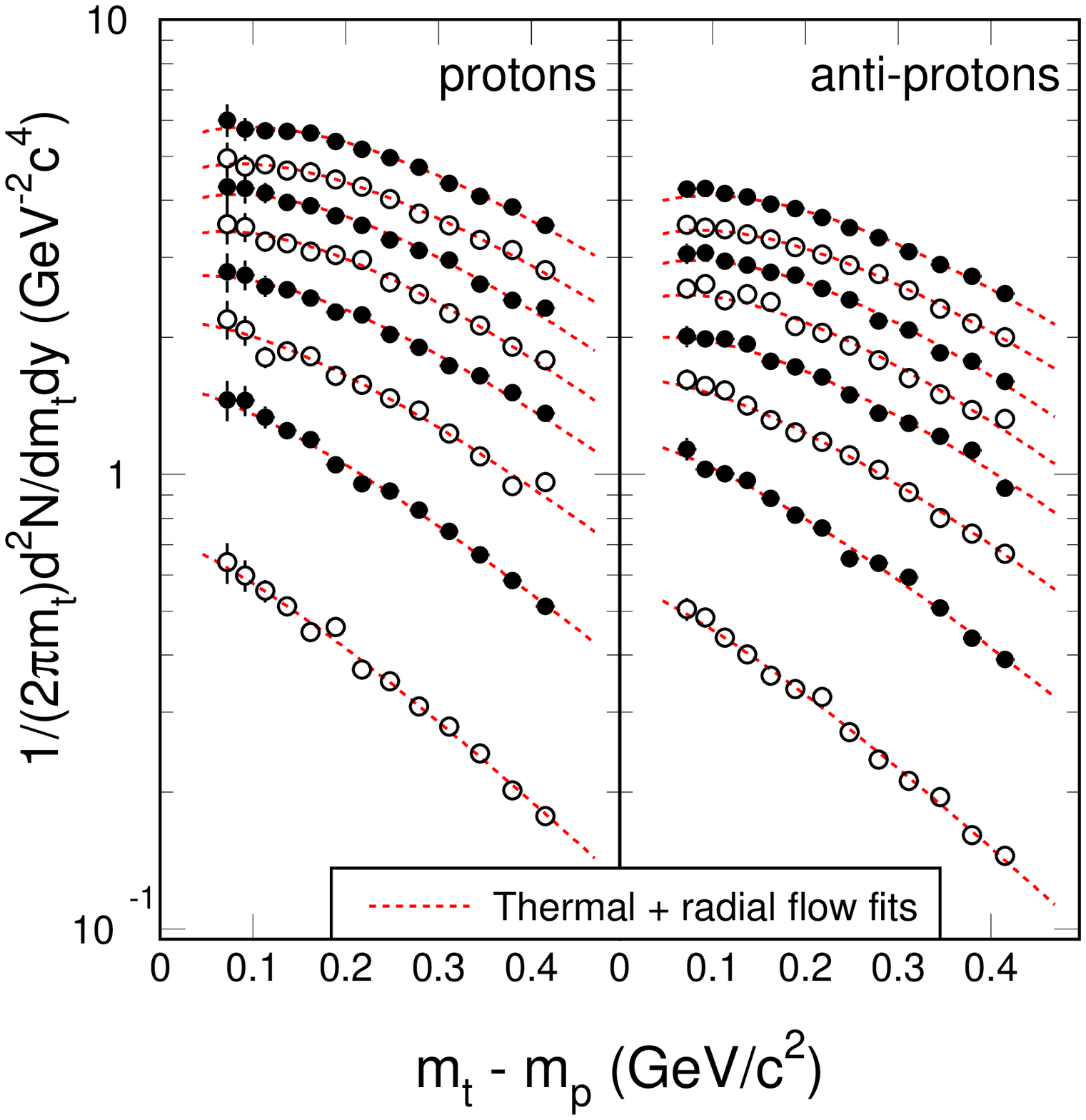}}
\caption{Mid-rapidity ($|y|\le0.5$) proton (left column) and
  anti-proton (right column) transverse mass distributions for most
  peripheral (bottom) to most central (top) collisions.  The
  definitions of the centrality bins are listed in Table I. Relatively
  large systematic errors for protons in the low $m_t$ region are due
  to the background subtraction. Results from model fits are
  shown as dashed lines.}
\label{fig2}
\end{figure}

The mid-rapidity ($|y|\le0.5$) proton and anti-proton transverse mass
distributions for all 8 centrality bins are shown in Fig.~\ref{fig2}.
Here, the transverse mass $m_t$ is given by $m_t = \sqrt{p_t^2 +
  m_p^2}$, with $m_p$ the rest mass of the proton.  
The uncorrelated bin-to-bin systematic errors are estimated to be less than
7\%. 
It is evident that both proton (left panel) and anti-proton (right panel)
distributions become more convex from peripheral to central
collisions indicating an increase in transverse radial flow.
In order to extract $p_t$-integrated yields, $dN/dy$ and mean transverse momenta \mpt,
hydrodynamically motivated fits~\cite{uli93} were applied, assuming a thermal
source plus transverse radial flow. The fit parameters are
the temperature $T_{fo}$ at kinetic freeze-out and the transverse radial
flow velocity $\beta_s$ at the system surface.
A velocity profile $\beta_t(r) = \beta_s (r/R)^{0.5}$ was used, where
$R$ is the radius of the source.
These fits simultaneously describe experimental spectra of charged
pions~\cite{starpion}, kaons~\cite{starkaon}, protons and
anti-protons. The fit-results are shown as dashed lines in Fig.~\ref{fig2}. 
The description of the experimental data is remarkably good.
When strong collective flow develops, the transverse mass distributions for
heavy mass particles will not have the simple exponential shape at low
transverse mass. Therefore, the hydrodynamically motivated two parameter fits
become necessary \cite{na44flow}.  
The increase of \mpt\ with centrality is indeed reflected in the values of
the collective velocity parameter \mbeta, which increase from about
(0.42$\pm$0.10)$c$ to (0.56$\pm$0.05)$c$ from the most peripheral to the most central
collisions, respectively.

Note that in~\cite{starpbarp}, the anti-proton transverse momentum
distributions were fitted with a Gaussian function in $p_t$.
The difference between the model fit results
and Gaussian fits in $p_t$ are $< 6$\% and $< 10$\% for \mpt\ and
integrated yields $dN/dy$, respectively. Using other functions, i.e.
exponential in $m_t$ and a Boltzmann function, the systematic uncertainty
in $dN/dy$ due to extrapolation is estimated to be less than 20\%. Similarly,
the systematic uncertainty in \mpt\ is less than 6\%. 
The total systematic uncertainty in $dN/dy$ is less than 22\%, adding the contributions
due to extrapolation (20\%) and the track reconstruction efficiency (10\%) in
quadrature.
\begin{figure}[h]
\centering\mbox{
\includegraphics{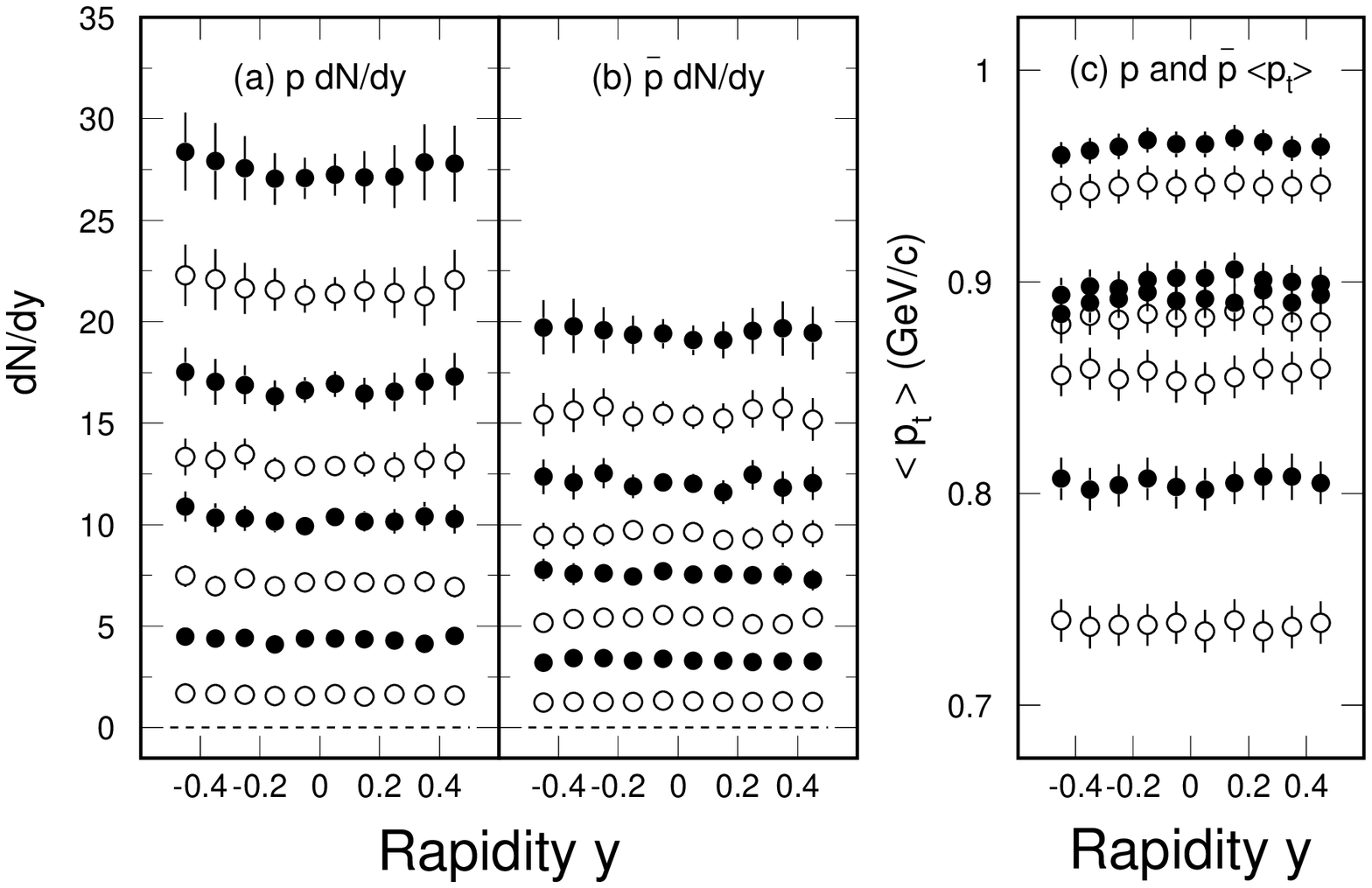}}
\caption{The rapidity distributions of protons~(a) and anti-protons~(b)  
  and the average transverse momentum \mpt\ (c),
  for most peripheral (bottom) to most central (top) collisions. 
  The bin-to-bin
  systematic errors due to PID contamination, were included in the
  plot. Overall systematic errors due to extrapolation into the
  $p_t$-range not covered by the experiment and the 
  uncertainty in the track reconstruction efficiency 
  are not shown in the figure.}
\label{fig3}
\end{figure}
The proton and anti-proton rapidity distributions are shown in
Fig.~\ref{fig3} (a) and (b) for different collision centralities.  In
the $p_t$-range not covered by this experiment, the yield was
extracted from the thermal model fit.  The results are shown in Table
I, which indicates that about 50\% of the integrated yield was measured
within the STAR TPC acceptance.  The bin-to-bin systematic errors, due
to background subtraction and PID contamination, are included in the
plot.  Since the shapes of the transverse mass distributions of protons
and anti-protons do not differ within statistical errors, the
extracted values of \mpt\ shown in Fig.~\ref{fig3}(c) are the average
of the two.  Within $|y|<0.5$, both values of \mpt\ and $dN/dy$ are
found to be uniform as a function of rapidity indicating that at RHIC
-- for the first time in heavy ion collisions -- a boost invariant
region of at least one unit of rapidity for all centrality bins has
developed.  An analysis of charged particle
ratios~\cite{BRAHMS_03} has demonstrated that at RHIC energies a
boost invariant region does not exist at $|y| > 1.5$. It will
be of interest to study the rapidity distributions of different mass
hadrons at RHIC.
%
%
\begin{figure}[th]
\centering\mbox{
\includegraphics{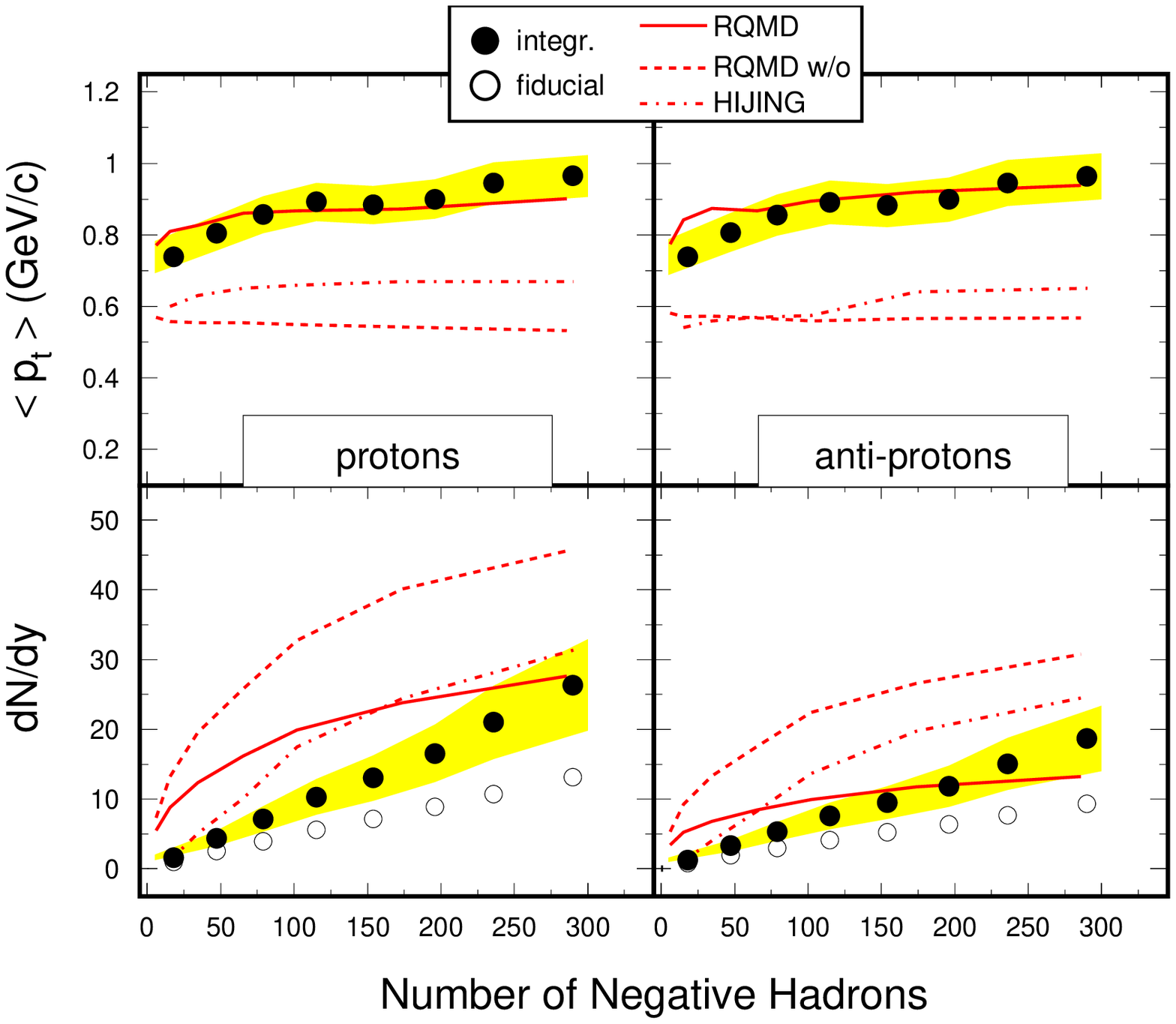}}
\caption{Mid-rapidity \mpt\ and $dN/dy$ of protons and anti-protons 
  as functions of the number of negatively charged hadrons. Open
  symbols are fiducial yields and filled ones are integrated yields.
  Systematic errors in the integrated yields are shown as shaded
  areas. Results from RQMD, RQMD with re-scattering switched off (w/o)
  and \hij\ are shown as solid-lines, dashed-lines and dashed-dotted
  lines respectively.  The experimental data and the results from RQMD
  and HIJING include feeddown from hyperon decay.}
\label{fig4}
\end{figure}

The top panels of Figure~\ref{fig4} show the \mpt\ within $|y|\le 0.5$
for protons (left) and anti-protons (right). The corresponding yields,
$dN/dy$ are shown in the bottom panels. The open symbols represent
fiducial yields and filled ones show the integrated yields.  The shaded
bands indicate the systematic uncertainties in extracting \mpt\ and
$dN/dy$.  Both values of \mpt\ and $dN/dy$ are in good agreement with
results from PHENIX \cite{phenixpikap}. 
Experimental results on the lambda(anti-lambda) yields~\cite{starlambda}, show that 
the contribution of feeddown from hyperon decays to the proton(anti-proton) 
yields is $\approx$40\%. 
The increase of \mpt\ vs.
centrality in the figure indicates the development of stronger
collective expansion in more central collisions.  Results from
calculations with RQMD~\cite{sorge95}, RQMD with re-scattering
switched off (w/o) and \hij~\cite{hijingpr,vance99} are represented by
solid, dashed, and dashed-dotted lines, respectively.  In the RQMD
model \cite{sorge95,d1} hadronic re-scattering has been implemented.
This leads to the agreement with measurements in the mean transverse
momentum. On the other hand, without the re-scattering, the
\hij\ model under-predicts the proton and anti-proton \mpt, especially
for central collisions.  Overall, the model calculations fail to
predict the experimental yields consistently throughout the whole
centrality range. Discrepancies between measured $\pbp$ ratios and
predictions from RQMD and HIJING have been reported by other
experiments~\cite{bramhs02, phobos_ratio}.

The bottom panels of Fig.~3 show that the observed
mid-rapidity ($|y|\le 0.5$) proton and anti-proton yields, $dN/dy$ are
proportional to the number of charged hadrons.  RQMD fails
to predict the centrality dependence of the anti-proton yield due to
the strong annihilation in hadronic re-scattering, especially in
central collisions.  Because of the annihilation, RQMD predicts a
change in the $\pbp$ ratio of almost a factor of two from peripheral
to central collisions, which is not consistent with
observations~\cite{starpbarp}. 

The results from RQMD reflect that within that model
there is strong annihilation among baryons, and that
large values of \mpt\ are built up from late hadronic
rescatterings. Based on RQMD, the annihilation of 
anti-protons created initially is expected to increase
from 20\% for perpiheral collisions, to 50\% for the
most central collisions. This is not consistent with
the trend in Fig.~3, which indicates the measured
proton and anti-proton yields increase approximately
linearly with the number of negatively charged hadrons.
This raises an important question. If, on the one hand
the increase in annihilation with centrality predicted 
by RQMD is correct, then the centrality dependence of
the initial baryon production must be much stronger
than the linear dependence observed in Fig.~3,
and the rough agreement between RQMD and the data for
anti-protons is fortuitous. If, on the other hand,
the agreement between RQMD and the linear dependence 
observed in Fig.~3 for anti-protons is correct, a 
possible explanation is that the anti-proton loss due 
to annihilation is smaller in central collisions than 
in peripheral collisions. This suggests the anti-protons 
may decouple from the surrounding matter early, and 
that the large experimental values of \mpt\ which are 
observed must arise from collective flow in the
early stage~\cite{cgc,starflow,starbb}. In order to distinguish this
possibility from other possible scenarios~\cite{rapp} and 
study possible early-stage partonic collectivity
at RHIC, systematic measurements of multi-strange
baryons, charmed mesons, and particle correlations
are necessary.

In summary, we have reported on the centrality dependence of proton
and anti-proton transverse mass and rapidity distributions from
$^{197}$Au+$^{197}$Au collisions at $\rts = 130$~GeV as measured by
the STAR experiment at RHIC.  The results reported here are from the
rapidity and transverse momentum range of $|y|<0.5$ and
0.35$<p_t<$1.00~GeV/$c$. For both protons and anti-protons, the
transverse mass distributions become more convex from
peripheral to central collisions indicating the enhancement of
collective expansion in more central collisions.  The rapidity
distributions and \mpt\ versus rapidity are 
found to be flat within $|y|<0.5$
suggesting a boost invariant region around mid-rapidity. 
The comparison of our
data to results from microscopic transport models suggests
that the observed collective expansion might have been dominantly
developed at the early stage of the collision.
%
\begin{table}
\caption{Mid-rapidity ($|y|<0.5$) proton and anti-proton results on
  \mpt\ and rapidity densities. 
The fiducial yield is measured within $0.35<p_t<1.00~{\rm GeV}/c$.
The errors are statistical. See text for discussions of 
systematic errors.}
\label{tab2}
\begin{center}
\begin{tabular}{r@{--}r|r@{$\pm$}l|r@{$\pm$}l|r@{$\pm$}l|r@{$\pm$}l|r@{$\pm$}l}
\hline \hline
\multicolumn{2}{c}{Cent.}           & 
\multicolumn{2}{c}{\mpt }     & 
\multicolumn{2}{c}{$dN_{p}/dy$}     & 
\multicolumn{2}{c}{$dN_{p}/dy$}     & 
\multicolumn{2}{c}{$dN_{\pbar}/dy$} & 
\multicolumn{2}{c}{$dN_{\pbar}/dy$} \\
\multicolumn{2}{c}{bin}         & 
\multicolumn{2}{c}{(MeV)}     & 
\multicolumn{2}{c}{(fiducial)}  & \multicolumn{2}{c}{(integrated)}     & 
\multicolumn{2}{c}{(fiducial)}  & \multicolumn{2}{c}{(integrated)}     \\
\hline
   58 & 85\% & 738 & 6 & 0.98 & 0.01 & 1.62 & 0.02  &  0.78 & 0.01  &   1.28 & 0.01  \\
   45 & 58\% & 805 & 6 & 2.51 & 0.02 & 4.36 & 0.05  &  1.91 & 0.02  &   3.31 & 0.03  \\
   34 & 45\% & 856 & 6 & 3.96 & 0.03 & 7.14 & 0.08  &  2.97 & 0.02  &   5.35 & 0.06  \\
   26 & 34\% & 892 & 6 & 5.55 & 0.04 & 10.29 & 0.10 &  4.08 & 0.03  &   7.56 & 0.07  \\
   18 & 26\% & 883 & 7 & 7.16 & 0.05 & 13.03 & 0.11 &  5.22 & 0.03  &   9.50 & 0.09  \\
   11 & 18\% & 900 & 8 & 8.92 & 0.06 & 16.53 & 0.14 &  6.40 & 0.04  &   11.85 & 0.10  \\
    6 & 11\% & 945 & 8 & 10.72 & 0.04 & 21.01 & 0.19 &  7.67 & 0.02  &  15.04 & 0.14  \\
    0 &  6\% & 965 & 7 & 13.17 & 0.04 & 26.37 & 0.23 &  9.35 & 0.02  &  18.72 & 0.16  \\
\hline \hline
\end{tabular}
\end{center}
\end{table}
\newline

We thank Drs. W. Busza, M. Gyulassy and V. Topor-Pop 
for exciting discussions. We wish to thank the RHIC Operations Group at
Brookhaven National Laboratory for their tremendous support and for
providing collisions for the experiment. This work was supported by
the Division of Nuclear Physics and the Division of High Energy
Physics of the Office of Science of the U.S. Department of Energy, the
United States National Science Foundation, the Bundesministerium f\"ur
Bildung und Forschung of Germany, the Institut National de la Physique
Nucleaire et de la Physique des Particules of France, the United
Kingdom Engineering and Physical Sciences Research Council, Fundacao
de Amparo a Pesquisa do Estado de Sao Paulo, Brazil, the Russian
Ministry of Science and Technology, the Ministry of Education of
China and the National Natural Science Foundation of China.

\end{document}